\documentclass{article}
\usepackage{siunitx}
\usepackage{graphicx}	
\usepackage{amsmath}
\usepackage{amssymb}
\usepackage{mathrsfs}
\usepackage{float}
\usepackage{color,soul}
\usepackage[title]{appendix}
\usepackage[a4paper, total={6in, 9in}]{geometry}
\usepackage{authblk}
\usepackage{multicol}
\usepackage{comment}
\usepackage[font={footnotesize}]{caption}
\usepackage{setspace}

\usepackage[numbers]{natbib}
\begin{document}

\title{Electron Emission Regimes of Planar Nano Vacuum Emitters}
\author[1]{Marco Turchetti \thanks{turchett@mit.edu}}
\author[2]{Yujia Yang \thanks{yujia.yang@epfl.ch}}
\author[1]{Mina R. Bionta}
\author[2]{Alberto Nardi}
\author[3]{Luca Daniel}
\author[1]{Karl K. Berggren}
\author[1]{Phillip D. Keathley \thanks{pdkeat2@mit.edu}}

\affil[1]{Research Laboratory of Electronics, Massachusetts Institute of Technology, Cambridge, MA 02139, USA }
\affil[2]{Swiss Federal Institute of Technology Lausanne (EPFL), CH-1015 Lausanne, Switzerland }

\date{\today}
\maketitle

\begin{abstract}
Recent advancements in nanofabrication have enabled the creation of vacuum electronic devices with nanoscale free space gaps.  These nanoelectronic devices promise the benefits of cold-field emission and transport through free-space, such as high nonlinearity and relative insensitivity to temperature and ionizing radiation, all the while drastically reducing the footprint, increasing the operating bandwidth and reducing the power consumption of each device. Furthermore, planarized vacuum nanoelectronics could easily be integrated at scale similar to typical micro and nanoscale semiconductor electronics. However, the interplay between different electron emission mechanisms from these devices are not well understood, and inconsistencies with pure Fowler-Nordheim emission have been noted by others. In this work, we systematically study the current-voltage characteristics of planar vacuum nano-diodes having few-nanometer radii of curvature and free-space gaps between the emitter and collector.  By investigating the current-voltage characteristics of nearly identical diodes fabricated from two different materials and under various environmental conditions, such as temperature and atmospheric pressure, we were able to clearly isolate three distinct emission regimes within a single device: Schottky, Fowler-Nordheim, and saturation.  Our work will enable robust and accurate modeling of vacuum nanoelectronics which will be critical for future applications requiring high-speed and low-power electronics capable of operation in extreme conditions. 
\end{abstract}

\section{Introduction}

Free-electron devices based on field emission played an important role in the early days of electronic systems development. In fact, vacuum tubes constituted the main building blocks of the first electronic computers, including Colossus\cite{randell1982}, used by the British to decipher German encrypted communication during WWII, and ENIAC\cite{hartree1946}, the first general purpose electronic computer, developed by the US Army to calculate ballistic trajectories. Vacuum tubes were then gradually substituted by semiconductor technology that could deliver faster switching times, lower power consumption, improved scalability and integrability, and did not require vacuum packaging.\cite{brinkman1997} The technology survived, but only in a few niche applications.\cite{gilmour2011,qiu2009,symons1998,barbour1998} However, in the last few decades, the advancement in nanofabrication techniques have allowed for the miniaturization of vacuum free-electron devices, which have started to regain interest due to their interesting properties when shrunk to the nanoscale.\cite{han2012}

Nano vacuum channel (NVC) electronics promise fast switching times, and low power-delay product with robust operation in harsh environments \cite{han2017}. Nanoscale vacuum channels allow for true ballistic transport with no phonon and charged impurity scattering, enabling higher electron velocities in the channel. Since these devices do not require vulnerable oxides and free-electrons are effectively insensitive to ionizing radiation and temperature fluctuations, these devices are attractive for applications in harsh environments such as space technology \cite{han2017, gaertner2012}. Field-emitter devices commonly use vertical geometries \cite{ding2000,ding2002,driskill1997,spindt1991} because with this approach it is easier to achieve sharp nanotips. However, it also makes them difficult to integrate with traditional electronics. On the other hand, planar NVC field-emitters could be easily incorporated into integrated circuits on a large scale. Moreover, thanks to their small size and low capacitance (down to tens of attofarads), they can be operated at petahertz-scale bandwidths, which makes them an ideal candidate for femtosecond electronics \cite{karnetzky2018} and other optoelectronic applications that require sub-optical-cycle response times \cite{schotz2019,rybka2016,yang2020,krausz2014}.

Despite these clear advantages, the underlying emission mechanisms of these devices is poorly understood. In literature, these devices are typically described using a pure Fowler-Nordheim tunneling emission model\cite{forbes2013}. While such a model can be used to fit the measured data, to do so requires the use of field enhancement factors of $\gamma > 100 \times$.  This stands in stark contrast to electromagnetic modeling of the tips which indicate only  modest field enhancement factors of $\gamma \sim 10\times$ for tips with nanometer-scale radii of curvature and gaps of few to tens of nanometers.  For instance, Nirantar et al. \cite{nirantar2018} had to assume a $\gamma = 590\times$ to accurately fit their results. Such discrepancies, exceeding more than one order of magnitude, were also noted by De Rose et al. \cite{de2020}, where they had to assume $\gamma = 133\times$ while their electromagnetic simulation would suggest $\gamma = 3.5\times$. This implies that Fowler-Nordheim tunneling is not physically consistent with the observed data, and implies some other emission physics was dominant. Understanding the dominant emission mechanisms involved and demonstrating how to reliably determine the proper regime of operation is critical if these device are going to be used to design and build electronic circuits that operate robustly in extreme environments.

In this work, we compared the emission characteristics of metallic (Au) and refractory (TiN) vacuum-channel bow-tie diodes having few-nm radii of curvature and sub-20-nm air/vacuum gaps. This comparison is of interest as the TiN devices are more resilient, allowing us to reach higher current densities from the emitters and thus transition to different emission regimes. 

While prior work has focused on three-terminal devices, we have chosen to focus on two-terminal diodes which allowed us to simplify the device geometry and focus on the underlying emission physics.  We showed that these vacuum nano-diodes can be operated reliably with turn-on voltages of $<$ 10 V and from nA to $\mu A$-level operating currents per device. We demonstrated repeatable behavior over many devices and over several scans per device. We analyzed the measured IV characteristics under variable temperature and atmospheric pressure to reveal the dominant mechanisms responsible for electron emission, and to rule out substrate conduction. In particular, we isolated threee distinct emission regimes from single devices for the first time: Schottky, Fowler-Nordheim field emission, and saturation. The transition between these regimes is still under scrutiny from a theoretical perspective, where some important effort has been done to build a model that would encompass all three.\cite{darr2020} We fitted our result with analytical models to better understand these behaviors ensuring that the field enhancement factor reasonably matches that from electromagnetic simulations.

\section{Results and Discussion}
\label{S:3}

\begin{figure}[h!]
	\centering
	\includegraphics[width = 1\linewidth]{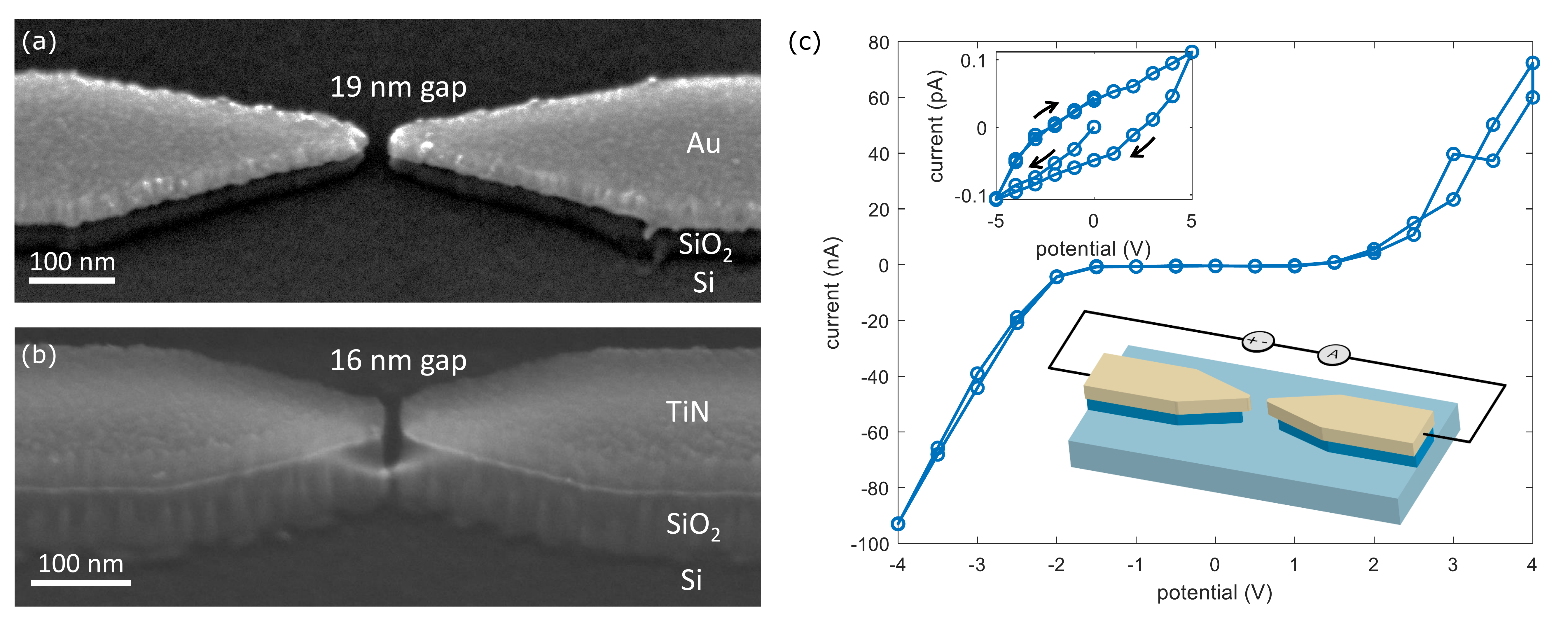}
	\caption{(a) SEM micrograph of a typical Au device. (b) SEM micrograph of a typical TiN device. We note that the thicknesses of the devices are different: the Au device is 25 nm thick while the TiN one is 50 nm thick. (c) Schematic of an IV measurement (bottom-right inset) and I-V curve of a metallic emitter with (main figure) and without (top-left inset) an undercut. As can be seen from the inset, without the undercut the current is very low and hysteresis is present. With the undercut, the current is much higher and the hysteresis disappears.}
	\label{Fig1}
\end{figure}

To analyze the emission behavior of these planar vacuum nanoemitters we fabricated both metallic (Au) and refractory (TiN) planar nano vacuum channel (pNVC) bow-tie diodes having ~10-20 nm vacuum gaps, using the procedures laid out in the methods section. Typical examples of the resulting structures are illustrated in the scanning electron microscope (SEM) micrographs shown in Fig. \ref{Fig1}a and Fig. \ref{Fig1}b. Particularly important was the implementation of an undercut in the fabrication process. In fact, preliminary testing showed that, without this undercut, the device often showed hysteretic behavior, which we attribute to the charging of the insulating layer underneath the devices. We found that introducing an undercut reliably eliminated this effect. An example of the effect of the undercut can be seen in Fig. \ref{Fig1}c where we measured an I-V curve of a metallic emitter with (main figure) and without (inset) the undercut. Moreover, an undercut allows us to ensure that the current we see is actually all due to emission in the vacuum gap and there is not a significant contribution due to surface conduction which can skew the analysis.  Additionally, we never imaged the devices before testing since the SEM electron beam causes the deposition of a carbon layer, and we observed that this can contribute to ohmic conduction. For imaging purposes, we always fabricated a twin device next to each device. 

\begin{figure}[h!]
	\centering
	\includegraphics[width = 1\linewidth]{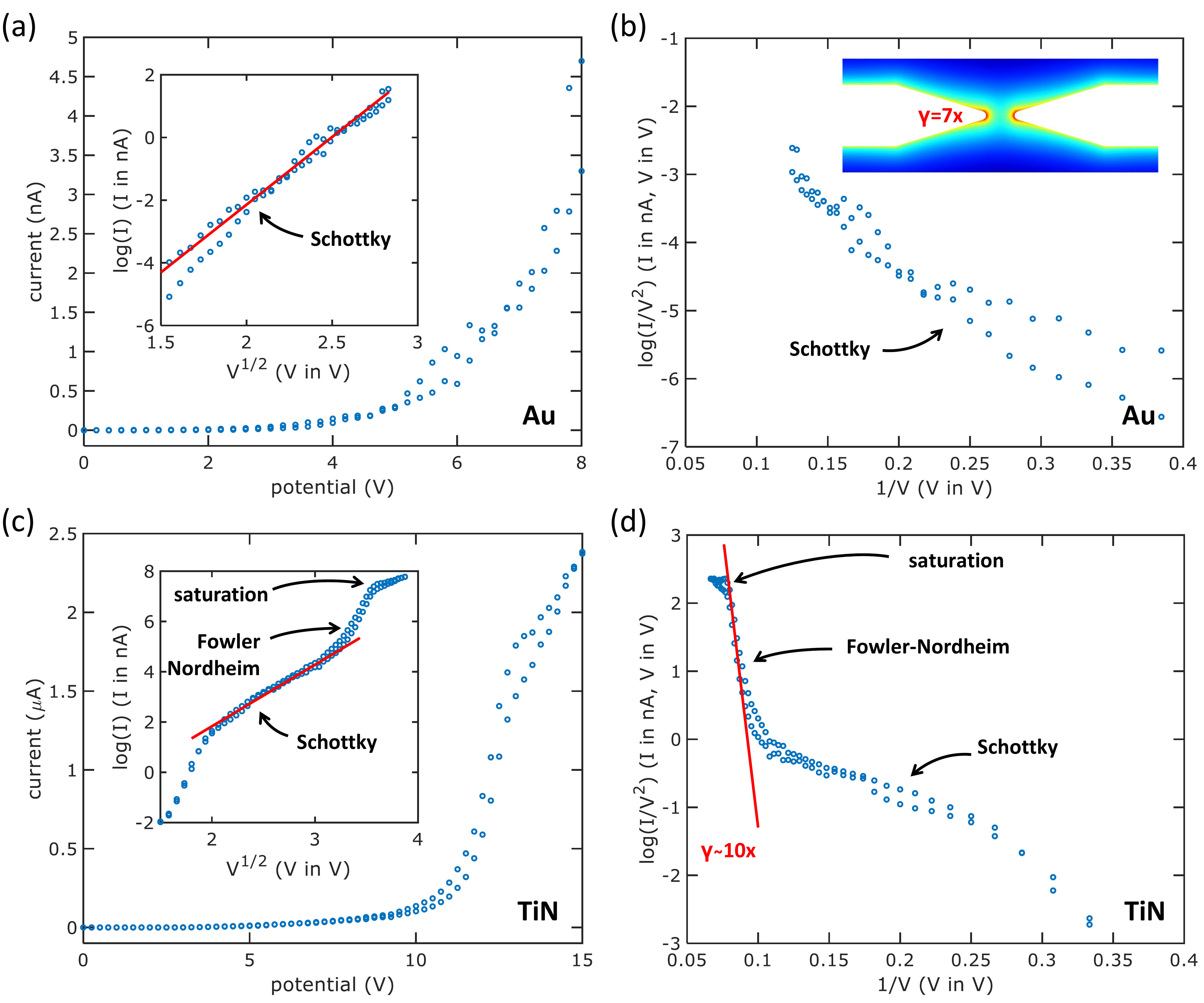}
	\caption{Typical behavior of an IV sweep of a Au and a TiN devices. (a) Au experimental results with inset $log(I)$ vs $V^{1/2}$ plot, which highlight Schottky behavior. We can see that the Au device exhibits Schottky behavior in this test, which can be identified by a linear dependence in the inset. (b) Au experimental results in $log(I/V^2)$ vs $1/V$ plot, which highlights Fowler-Nordheim behavior. Inset illustrates a 3D electromagnetic simulation showing the field enhancement factor ($\gamma$) in the region around the two tips, highlighting a $\gamma = 7\times$ at the edge of the tip. From these data, to fit a Fowler-Nordheim emission we would need to assume a $\gamma = 50\times$, which is inconsistent with simulation.(c) TiN device experimental results with inset Schottky plot. In this plot, we can identify a Schottky regime which manifests as a linear dependence in the inset. This is followed by a superlinear regime and then a saturation.  (d) TiN experimental results in Fowler-Nordheim plot. Here, we can identify that the superlinear regime that was visible in the Schottky plot is indeed driven by Fowler-Nordheim emission. In fact, this region can be fitted with a FN model, assuming a $\gamma = 10\times$, which is consistent with the simulation. These tests were performed at $10^{-6}$ mbar.}
	\label{Fig3}
\end{figure}

After the fabrication, we proceeded with analyzing the I-V response of tens of devices fabricated to investigate the underlying emission mechanisms. The testing was performed in a vacuum chamber with pressure of $10^{-6}$ mbar using a Keysight B2912A SMU. Fig. \ref{Fig3}a shows an example of experimental data for a gold bowtie nano emitters having a gap of $<20$ nm. The device current exhibited an exponential behavior with respect to the applied field with a turn on voltage of approximately 5V which is consistent with the Au work function of 5.1eV. 

The inset shows the same curve plotted in a $log(I)$ vs $V^{1/2}$ plot, which is useful for identifying emission in Schottky regime. The Schottky emission regime applies to field-enhanced thermionic emission. In the Schottky emission regime, the applied field reduces the work function barrier height, and enhances the thermionic emission. Schottky emission can be modeled as \cite{tomer2015}:

\begin{equation}
I \propto T^2 \mathrm{exp} \left( \frac{q}{2k_B T}\sqrt{ \frac{q\gamma V}{d\pi\epsilon_0}}\right) \mbox{,}
\end{equation}
where $\gamma$ is the field enhancement factor, $T$ is the temperature, $\epsilon_0$ is the vacuum permittivity, $q$ is the electron charge, $k_B$ is the Boltzmann constant,  $V$ is the potential, and $d$ is the gap between the tips.  Schottky emission therefore would appear linear with a positive slope when plotting $log(I)$ vs $V^{1/2}$.  As such, we refer to these plots as ``Schottky plots'' for simplicity throughout the remainder of this work.  We can see that the tested gold device exhibits a linear characteristic in this plot over the entire range of bias voltages tested, indicating that Schottky emission could be dominant. 

To ensure our interpretation is correct, we also considered emission due to Fowler-Nordheim tunneling. The Fowler-Nordheim regime applies to cold-field-emission where the electrons tunnel through the work-function barrier from the Fermi surface of the material, and is the most commonly used theory for modeling field-induced electron emission in literature. Fowler-Nordheim emission can be modeled analytically as:

\begin{equation}
I \propto \phi^{-1}\left(\gamma \frac{V}{d}\right)^{2} \mathrm{exp} \left( -b\frac{d \phi^{3/2}}{\gamma V}  v(y) \right) \mbox{,}
\end{equation}

where $v(y) = 1 - y^2 - y^2\mathrm{log}(y)/3 $, $y=2\sqrt{\frac{e^2 \gamma V}{16d\pi\epsilon_0}}\frac{1}{\phi}$, $\phi$ work function and $b = 6.83$ $eV^{-3/2} Vnm^{-1}$.
For the FN fit we used a more complete version of this formula, which can be found in Kyritsakis e al. \cite{kyritsakis2015}



Fig. \ref{Fig3}b illustrates the same data as in \ref{Fig3}a, but now plotted in a Fowler-Nordheim plot ($log(I/V^2)$ vs $1/V$).  In such a plot, FN emission would appear linear with negative slope. While a linear trend with negative slope does indicate field-emission as likely, it unfortunately does not guarantee that Fowler-Nordheim-like tunneling is truly dominant.  We note that Schottky emission can also appear quasi-linear when plotted in this fashion over a given bias voltage range.  Indeed, this is the case with the data from our gold devices.  
However, when we then fit the curve with a FN model we have to assume $\gamma = 50\times$, which is not consistent with our electromagnetic simulation results.  Electromagnetic simulations consistently predict $\gamma \sim 10\times$. An example of such a simulation illustrating $\gamma$ around the two tips is shown in the inset of Fig. \ref{Fig3}b, which shows a peak of $\gamma \approx 7\times$. Taken together with the Schottky plot, this provides strong evidence that across the tested bias range Schottky emission dominated from the gold devices. This is further confirmed with temperature testing which is described below. Unfortunately, we were not able to run the device at higher potentials to determine if there was a point where Fowler-Nordheim emission appears dominant due to degradation of the devices. This may be due to reshaping of the tips due to high current density or modification of the work function due to current-assisted adsorption on the tips.  


To investigate what impact the emitter material might have on the emission properties, we then performed similar testing on the TiN devices. Thanks to their more resilient nature and a thicker oxide, we were able to run the TiN devices at higher fields and current densities, which allowed us to see transitions between different emission regimes as can be seen in Figs. \ref{Fig3}c and d.  At low voltage (approximately from 4 V to 10 V) we observe what appears to be Schottky emission behavior (see linear response over this range in the Schottky plot shown in the inset of Fig. \ref{Fig3}c).  Unlike the gold devices, at higher potentials (between 10V and 13V) we can see a transition away from Schottky emission where the emission grows at an even higher exponential rate. Fig. \ref{Fig3}d illustrates the same data in a Fowler-Nordheim plot. In this case, we note that the slope is considerably steeper than the slope for Au for bias voltages between 10 to 13 V. Indeed, the data in this region can be fitted using the aforementioned FN model, which predicts $\gamma \approx 10\times$, which is physically consistent with the $\gamma \approx 7\times$ obtained from our electromagnetic simulations. Finally, above 13V the current reaches a saturation region. We note that this behavior is repeatable in forward and backward scans so it is not due to damage of the devices.  We attribute this saturation behavior to Child-Langmuir regime\cite{umstattd2005,lau1994}, where the emission current is limited by the space-charge-effect.

\begin{figure}[h!]
	\centering
	\includegraphics[width = 1\linewidth]{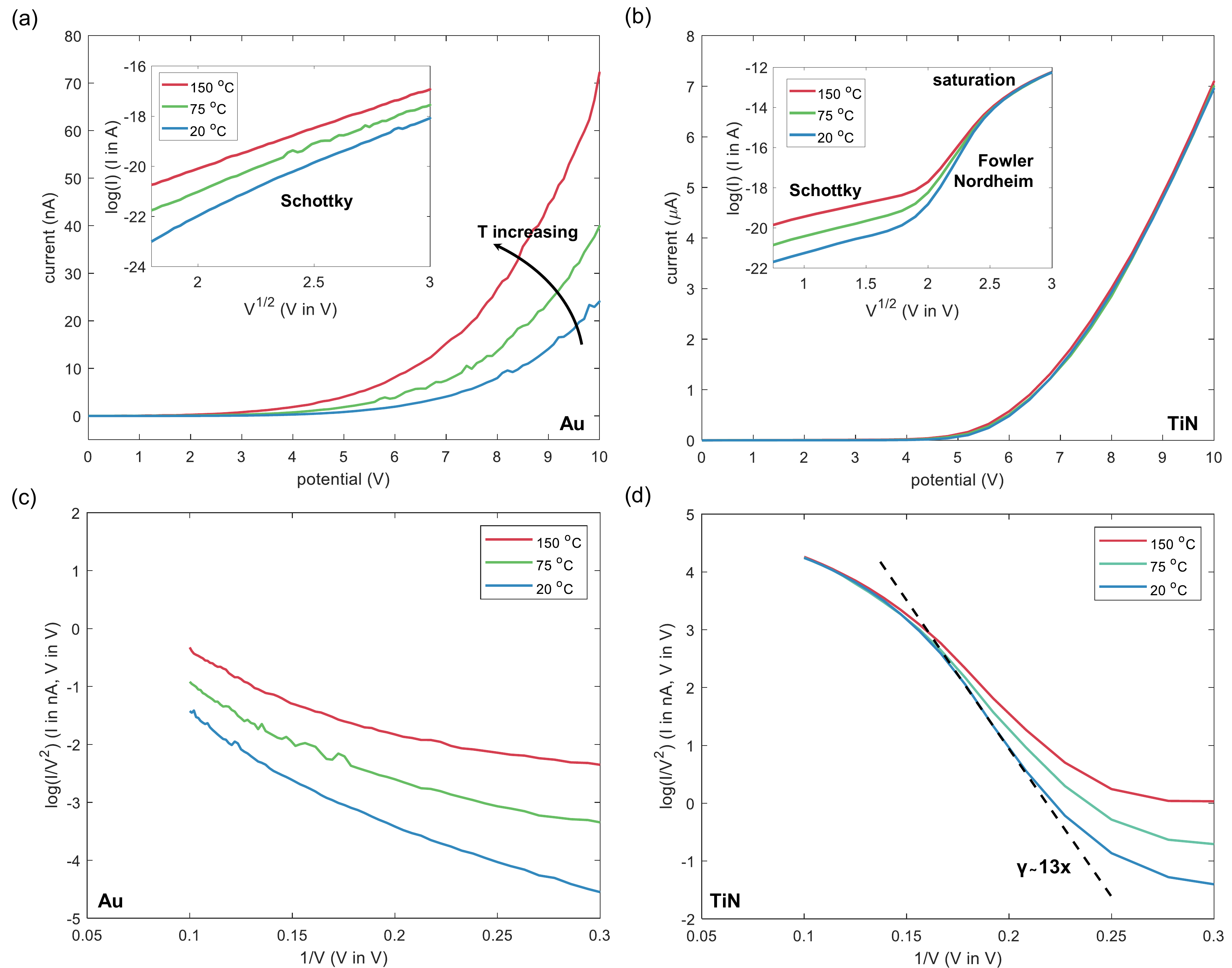}
	\caption{Temperature dependence of Au (a) and TiN (b) devices. The devices' IV characteristics are recorded varying the device temperature using a heater placed in thermal contact with the sample. The insets illustrate the same data plotted in a $log(I)$ vs $V^{1/2}$ axes. The Au devices exhibit Schottky behavior which which manifests as parallel traces in this plot. On the other hand TiN devices exhibit all three regimes (Schottky, Fowler-Nordheim and saturation). While Schottky regime shows temperature dependence similarly to the Au devices, Fowler-Nordheim does not which is consistent with the model. Saturation regime also does not show any temperature dependence. It is worth noticing that in this case the TiN device enter in Fowler-Nordheim and then saturation at a lower voltage than the Au device. This can be due to different gap sizes or a sharper tip: both parameters that can vary with slightly different fabrication conditions. The same Au and TiN data is plotted in a FN plot in (c) and (d) respectively. }
	\label{Fig5}
\end{figure}

To further investigate our findings, we next tested the temperature dependence of the devices' IV response. Such tests should clearly differentiate between Schottky and Fowler-Nordheim emission as Schottky emkission depends strongly on temperature, while Fowler-Nordheim emission does not. The temperature tests were done by placing a heating stage inside the vacuum chamber in thermal contact with the sample. The results of this test are shown in Fig. \ref{Fig5} for both Au and TiN devices. The Au devices (Fig. \ref{Fig5}a) exhibits a clear temperature dependence over the entire range of applied voltages.  Each scan appears in the Schottky plot as a series of spaced, roughly linear traces consistent with the expected behavior for Schottky emission. In this regime the temperature provides the necessary energy to overcome the barrier set by the work function and applied bias voltage.

On the other hand, for the TiN devices, we can clearly see all three regimes. In the Schottky regime, which dominates at the lowest voltages, there is a temperature dependence which manifests as vertically-spaced parallel traces in the Schottky plot in the inset of Fig. \ref{Fig5}b. This dependence gradually shrinks and then disappears when Fowler-Nordheim  tunneling begins to dominate at higher voltages. This reduction in temperature dependence is consistent with cold field emission, where tunneling from near the Fermi level dominates. Finally, we also observe no temperature dependence in the saturation regime. This is consistent with Child-Langmuir space-charce saturation which is a charge-density-induced limitation that does not depend on temperature or material properties.

When the same data are plotted on a FN graph (Fig. \ref{Fig5}c,d), we can see that the TiN device reaches Fowler-Nordeim regime approaching the $\gamma \approx 13\times$ curve (fitted using the $20 ^{\circ}$C data) and then saturates similarly to what is predicted by Lau et al. \cite{lau1994} for a Fowler-Nordheim to Child-Langmuir transition .

\begin{figure}[h!]
	\centering
	\includegraphics[width = 1\linewidth]{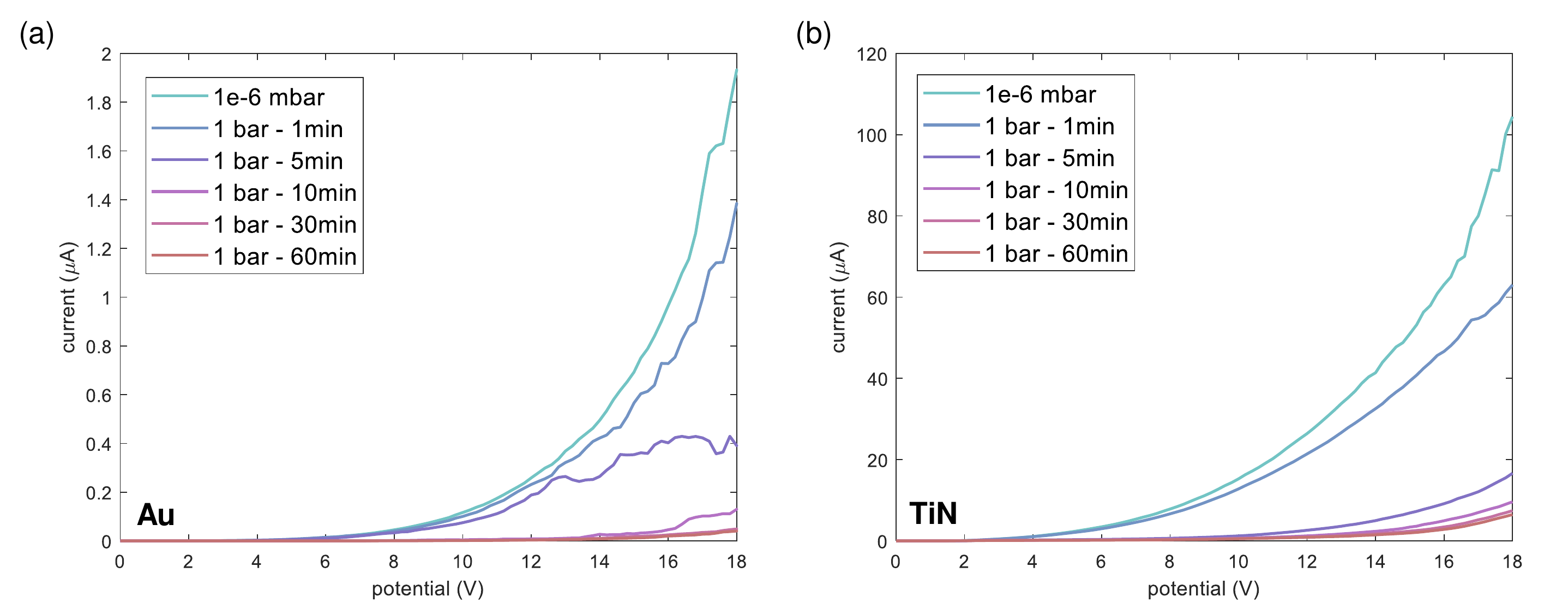}
	\caption{Pressure dependence of Au (a) and TiN (b) devices. The devices are first tested in a vacuum chamber with a $10^{-6}$ mbar vacuum. Then the chamber is vented with air and a series of consecutive traces are recorded at different time intervals: 1, 5, 10, 30 and 60 min. The drop in current is due to adsorption on the tip surface and shows that the conduction is indeed in the vacuum channel with no significant contribution due to substrate conduction, which would not be affected by the pressure.}
	\label{Fig6}
\end{figure}

We also investigated the influence of ambient air pressure on the devices. To do this, we initially tested the devices at 1e-6 mbar and then we vented the chamber and recorded the IV characteristic at different intervals after exposure to ambient air.  The sub-20nm gaps ensure that even in ambient pressure the conduction through the emitter to collector gap happens in an effective vacuum, not mediated by the gas. This is a because the air molecules mean free path at atmospheric pressure is larger than the gap size. However, as can be seen in Fig. \ref{Fig6}, when exposed to the atmosphere, these devices, nonetheless experience a strong reduction in current. The current stabilizes after about an hour to a a level much lower than that in vacuum conditions, both for the metallic and refractory devices. We attribute this effect to adsorption of molecules (e.g. water) on the tips which modify the work function at the emission surfaces. In the case of TiN devices oxidation might also play a role. In both cases, the vacuum emission characteristics can be fully recovered after a few burn-in cycles (i.e. running a few IV sweeps that clean the emission surface from absorbed molecules) once they are placed back in vacuum. It is also noteworthy that the current drop experienced by the refractory devices when exposed to atmospheric pressure is less dramatic than that experienced by the metallic devices. Indeed, the refractory device in Fig. \ref{Fig6} experience a current drop of $16\times$, while the gold device experience a current drop of $50\times$.  These findings further confirm the ballistic nature of the electron emission and transport through the vacuum channel as contribution to the conduction due to leakage through the substrate would not exhibit such strong sensitivity to air exposure.

\section{Conclusion}
\label{S:4}

We investigated the emission physics of planar NVC diodes. To do so, we analyzed the IV characteristics of two different emitter materials, Au and TiN, under varying temperatures and atmospheric conditions.  While past work had primarily used Fowler-Nordheim tunneling to model the emission physics of such devices, a large discrepancy was found between the fitted electric field enhancement factor and that expected from electromagnetic modeling.  Upon closer inspection in this work, we found that Schottky emission tends to dominate at lower applied bias values. In particular, we found that for the Au devices we were only able to observe temperature-dependent Schottky emission before the onset of damage. Instead, for TiN, thanks to the possibility of exploring higher potentials given their higher physical robustness, we were able to identify the transition from Shottky to Fowler-Nordheim tunneling before a final transition to saturation. We ascribe the saturation regime to the Child-Langmuir space charge limitation. Depending on the device requirements (e.g. low voltage or high transconductance) devices could be designed to operate in different regimes. These findings mark an important step toward the development of accurate models necessary for the designa and realization of high-speed\cite{karnetzky2018}, robust\cite{bhattacharya2021} and radiation-resistant\cite{han2017} vacuum nanoelectronics.

Finally, the pressure analysis revealed a large reduction in the emission rate and increase in the turn-on voltage of the devices due to a combination of oxidation and absorption of molecules on the surface. While this sensitivity verifies tunneling emission and transport through free-space dominates compared to substrate leakage, it unfortunately indicates that such devices should be properly packaged despite the reduced free-space channel width. However, we emphasize that this degradation is reversible once vacuum was restored.

\section{Methods}
\label{S:2}

We explored two different materials for these devices: Au and TiN. Therefore, we developed two different fabrication techniques. The patterning of the Au devices is achieved through a lift-off process. Instead, the patterning of the TiN is achieved through etching using an hard mask. In the following we are going to illustrate the main steps of these fabrication. 

\subsection{Gold structures}
\begin{figure}[h!]
	\centering
	\includegraphics[width = 0.8\linewidth]{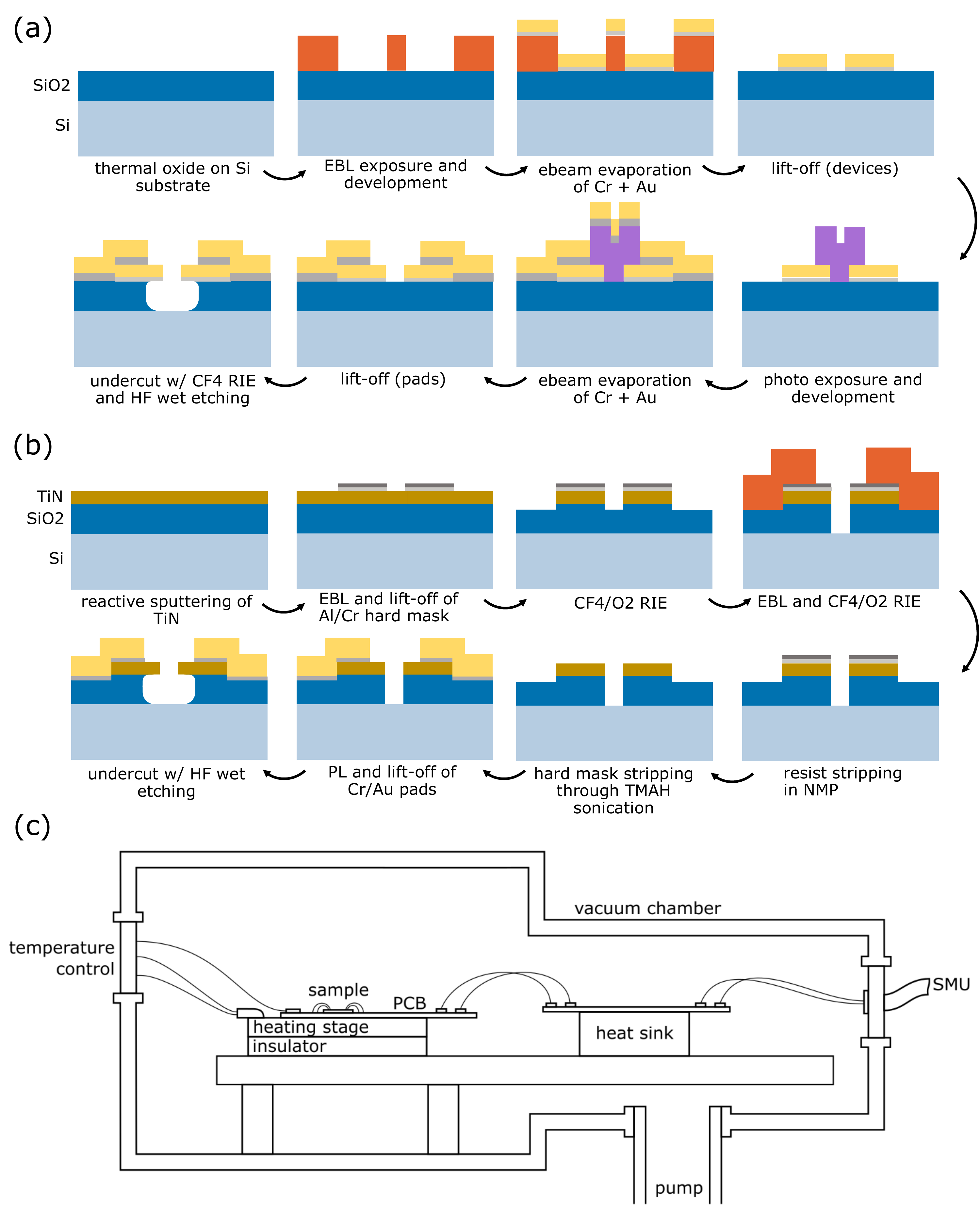}
	\caption{(a)Au nanofabrication process. (b) TiN nanofabrication process. (c) Measurement setup.}
	\label{Fig2}
\end{figure}

We developed this process for Au devices but, in general, it can be extended to the patterning of any metallic material that can be e-beam evaporated. Fig \ref{Fig2}a is a graphical illustration of the different steps of the process:

\begin{enumerate}
	\item EBL patterning of PMMA A2 resist on a thermal oxide on Si substrate: this step is performed to pattern the devices;
	\item resist development and ebeam evaporation of 5 nm Cr and 20 nm Au;
	\item lift off in heated NMP;
	\item photolithography of a bilayer PMGI+S1813 resist: this step is performed to pattern the pads for the electrical connections;
	\item resist development and ebeam evaporation of 30 nm Cr and 150 nm Au
	\item lift off in NMP
	\item CF4 RIE and 40s of (9:1) DI:BOE HF: step that creates an undercut at the tip
\end{enumerate}
We did some preliminary test without the last step, but we observed a low current and hysteretic IV characteristics of the devices. We concluded that such effect was caused by emitted electrons that gets trapped in the oxide creating a repulsive potential which prevented further electrons to be emitted. Once the etching step used to create the undercut was introduced the effect disappeared.
An example of a completed structure done with this process is shown in Fig \ref{Fig1}a. 

\subsection{TiN structures}
We developed this process for TiN devices but, in general, it can be adapted to the patterning of many hard materials, modifying the first etching step chemistry. Fig \ref{Fig2}b is a graphical illustration of the different steps of the process:
\begin{enumerate}
	\item reactive sputtering of TiN 
	\item EBL patterning of PMMA A2 resist on a thermal oxide on Si substrate: this step is performed to pattern the hard mask;
	\item resist development and e-beam evaporation of bilayer hard mask (15 nm Al and 15 nm Cr). We determined that Cr hard mask gives the best results in terms of sidewalls steepness but it is hard to remove. Using a bilayer mask allows us to exploit the Cr benefits and then remove it with an Al lift-off;
	\item lift off in heated NMP;
	\item CF4/O2 RIE etching: this step is performed to pattern the devices;
	\item EBL patterning of ZEP502A resist: this step is performed to pattern a region that will undergo the etching for creating the undercut;
	\item CF4/O2 RIE etching;
	\item ZEP502A resist stripping in NMP;
	\item mask lift off through sonication in TMAH;
	\item photolithography of a bilayer PMGI+S1813 resist: this step is performed to pattern the pads for the electrical connections;
	\item resist development and ebeam evaporation of 30 nm Cr and 150 nm Au
	\item lift off in NMP
	\item 70s of (9:1) DI:BOE HF: step that creates an undercut at the tip
\end{enumerate}
An example of a completed structure done with this process is shown in Fig. \ref{Fig1}b.

Fig. \ref{Fig2}c illustrates the schematic of the testing apparatus.

\bibliography{references}

\section*{Acknowledgements}
This work was supported by the Air Force Office of Scientific Research under award numbers FA9550-19-1-0065, and FA9550-18-1-0436. 
We would also like to thank the internal reviewers Emma Batson and Stewart Koppell of the Quantum Nanostructures and Nanofabrication group at MIT.
\end{document}